\documentstyle[aps,preprint,epsf]{revtex}

\begin{document}
\title{Influence of surfactants on the structure of titanium oxide gels :
experiments and simulations}
\author{F. Molino$^{\dag}$ }
\author{J. M. Barthez$^{\dag \ddag}$}
\author{A. Ayral$^{\ddag}$ , C. Guizard$^{\ddag}$, R. Jullien*, J.
Marignan$^{\dag}$}

\address{$^{\dag}$ Groupe Dynamique des Phases Condens\'ees, U.A. 233 C.N.R.S.,
Universit\'e Montpellier II, 34095 Montpellier, Cedex 05, France}
\address{$^{\ddag}$ Laboratoire des Mat\'eriaux et Proc\'ed\'es Membranaires,
U.M.R. 9987 C.N.R.S., E.N.S.C.M., 8, Rue de l'Ecole Normale, 34053 Montpellier,
Cedex 01, France}
\address{*Laboratoire de Science des Mat\'eriaux Vitreux, U.A. 1119 C.N.R.S.,
Universite  Montpellier II, 34095 Montpellier, Cedex 05, France}

\maketitle
\begin{abstract}
We report here on experimental and numerical studies of the influence of
surfactants on mineral gel synthesis. The modification of the gel structure
when the ratios water-precursor and water-surfactant vary is brought to the
fore by fractal dimension measures. A property of {\em polydispersity of the
initial hydrolysis} is proposed to explain these results, and is successfuly
tested through numerical experiments of three dimensional chemically limited
aggregation.
\end{abstract}

\newpage
\section{INTRODUCTION}
The irreversible aggregation of molecules in a solution, leading to the
formation of a three-dimensional (3D) network or {\em gel}, has been the focus
of a wide range of studies, from the more theoretical to the more
application-oriented ones\cite{1p}, \cite{5}.
The structure of the gel is a well-known case of {\em fractal mass
distribution} \cite{7}.\\
The polymerization of a mineral monomer in an organic solvent is an example of
this sol-gel process. More specifically, a very studied class of monomers is
the {\em metal alcoxides} one, which elements consist of a metal atom (as
silicon(Si) or titanium (Ti)) surrounded by alcoxy (alcohol) groups. In an
organic solvent (decane, cyclohexane) and in the presence of water, they
readily react in a two-step process leading to the fractal polymer.\\
In the first step, {\em hydrolysis}, the interaction of monomers with water
leads to the substitution of {\em some} alcoxy groups with short -OH (hydroxy)
radicals (the number of groups modified depending at least of the total amount
of water available):
$$
Ti(O R)_4 + n H_2O \rightarrow (HO)_n Ti (OR)_{4-n} +(4-n) ROH.
$$
Then the hydrolyzed monomers stick together throught -OH HO- (oxolation) or -OH
RO- (alcoxolation) reactions,  to form the inorganic backbone of the fractal
polymer (this step is the {\em polycondensation}):
$$
- Ti - OH + HO - Ti - \rightarrow H_2O + - Ti - O - Ti -  .
$$
Due to the presence of four potentially active sites, one obtains a branched
polymer and the gelation of the structure. \\
However, for some very reactive alkoxide precursors (as in our case
tetraisopropyl orthotitanate [$Ti(O^iPr)_4$]), the reaction is so fast, even
for very small amounts of water, that  the precipitation of the solution can be
avoided only by introducing amphiphile molecules which control the sol-gel
transition by interacting with water or hydroxyl radicals. Only this trapping
effect obtained from the introduction of the surfactant allows the formation of
the gel by slowing down the hydrolysis, as has been recently investigated
\cite{1} \cite{2} \cite{3}.\\
Rather surprisingly, considering the stability of the caracteristic {\em
fractal dimension} of the gel with variations of many parameters, the
amphiphilic molecules not only control the kinetics of the reaction, but induce
structural variations in the final polymer. In the case where a gel could be
synthetized, we have studied experimentally the influence on the process of the
water-surfactant and water-precursor ratios. More specifically, we report here
of the possibility of understanding some of the observed variations of the gel
structure from  the following hypothesis: in reactions with very small amounts
of water, not all the sites on the monomers  will be hydrolyzed and  {\em the
distribution among the monomers of the hydroxyl (-OH) radicals created in the
hydrolysis on the precursor monomers could depend, for a given amount of water,
on the ${\displaystyle [H_2O]/[Surf.] }$ ratio }. Numerical experiments which
have been done to test this idea are described in this paper, and explain some
evolutions of the fractal dimension of the final gel.\\
\section{EXPERIMENTAL RESULTS}
The gels have been synthetized in two different solvents (cyclohexane and
decane) from a $Ti(O^iPr)_4$ monomer. For each solvent, two solutions were
prepared, the first (S1) containing the surfactant (Triton X-35) and the water,
and the second (S2) the precusor; S1 was then tipped into S2. The
characteristic parameters of the solution were the  $T_i(O_iPr)$ molality,
always fixed to $0.1$  $mol. kg^{-1}$, and the molar ratios:
$$
n = \frac{[H_2O]}{[Surf.]},\mbox{\hspace{5mm}}
h = \frac{[H_2O]}{[Ti(O^iPr)_4]}.
$$
These diluted sols exhibit long gelation times and produce gels with a large
fractality range \cite{3}. The polymeric gelation could be obtained for $h$
between 2 and 3, and $n$ between 1 and 2. The structural characteristic of the
final gel we were interested in was the {\em fractal dimension}, which was
measured using ultra small angle x-ray scattering \cite{3p}. One can see in
fig. ~\ref{fig1} the evolution with $h$  of the fractal dimension, for two
values of $n$.\\
As previously said, the kinetic of the sol-gel process can be decomposed in two
steps: hydrolysis of the isopropoxy group, and polycondensation of the hydroxyl
radicals. Let's go into some details.
On the basis of the partial-charge model \cite{4}, alcoxolation (condensation
between $OH$ and $OR$ groups) should be the favored condensation reaction
between partially hydrolyzed, coordinatively saturated titanate precursors. As
underlined by Brinker and Scherer \cite{5}, the same trend is predicted for
silicate but $^{29}Si$ NMR investigations \cite{6}  have shown that oxolation
(condensation between $OH$ groups) is the favored reaction of condensation
between partially hydrolyzed silicon alkoxides. Perhaps solvent interactions
that are not accounted for by the partial-charge model influence the quality of
the leaving group \cite{5}. In our case, the assumption that oxolation is the
favored reaction can also be justified taking into account the stronger
lyophobicity of the hydroxyl groups in the inverse micellar medium.\\
An NMR study of the isopropanol amount in the solution \cite{3} led us to the
conclusion of an almost instantaneous hydrolysis compared with the gelation
time. Therefore one can hope to study separately the effects of the surfactant
presence on the two reaction steps, at least if no important "second
hydrolysis" occurs, due to the water created during the condensation.  Our
modelization assumption concerns primarily the way the surfactants could
control the hydrolysis. In the solution, the water is trapped on the
hydrophilic heads of the amphiphils (with a limit of four molecules of $H_2O$
on every surfactant), which organize themselves in inverse micellar aggregates
in the hydrophobic solvent. Quasielastic light scattering measurements \cite{1}
give us a hydrodynamic radius for these structures between 3 nm for $n=1$ and 4
nm for $n=2$, the surfactant itself having an average extension of 1.5 nm. One
can estimate from these radii that the number of pseudo-micelles for one
alkoxide  is $3.5$ times larger for $n=1$ than for $n=2$. In order to be
hydrolysed, the precursor molecules must penetrate into these aggregates. This
was the motivation for introducing our "hydrolysis polydispersity" hypothesis:
the number of hydroxyl radicals created on the alkoxide will vary between 1 and
4, with a distribution dependent on $n$, due to the variation with this
parameter of the available water in the micellar structure.\\
\section{THE MODEL AND THE SIMULATION}
Let us describe the numerical experiments performed in order to test this
assumption. In our three-dimensional algorithm, every particle has six
potentially active sites (cubic symmetry). Taking into account the titanium
tetravalence, we hydrolyze at most four of these sites, always situated in the
same randomly selected plane. The hydrolysis ratio parameter $h$ is represented
by a scalar $q$ varying between $0.5$ and $0.75$ for  $h$ between $2$ and $3$ :
for N particle samples, the hydrolysis stops when $4 N q$ sites have been
hydrolyzed. The implicit hypothesis in taking $q=h/4$ is that all the water
contained in the solution has been used in the hydrolysis (a conclusion we
inferred from our NMR studies). It is far more difficult to introduce $n$ in
the algorithm, for many hypotheses on the molecular behavior are involved in
the prediction of its influence on the hydrolysis. We have chosen a parameter
$p$ in $[0,1]$ to govern directly the polydispersity in hydrolysis of the
sample. The idea is as follows : during every round a monomer is randomly
chosen and a random number $z$ generated. Depending on wether $z>p$, $p^2<z<p$,
 $p^3<z<p^2$, or $z<p^3$, 1, 2, 3, or 4 sites are {\em simultaneously
hydrolyzed}. Then a second particle is selected, {\em among the ones which have
not yet been hydrolyzed during this round}. When every particle has been
selected exactly once for hydrolysis in the round, a next round opens, and so
on until the $4 N q$ water molecules available have been used.\\
The algorithmic interest of this procedure is to make the hydrolysis
polydispersity vary strongly with $p$, which increases all eventual
polydispersity effects. This does not happen, for example, if at every step the
hydrolyzed particle is put again in the sample from which the next one is
selected. But what about the  physical point of view? Concerning the power law
in $p$ for the choice of the number of hydroxyl radicals created, an {\em
independence} of the different hydrolyses happening with the same probability
when the precursor is in the micellar structure is implicit. It may appear more
arbitrary to hydrolyse every particle exactely once in ever round. As
previously remarked, this is essentially in order to make the hydrolysis
polydispersity vary more strongly in the algorithm.\\
Concerning the {\em aggregation dynamics}, we have chosen a {\em
chemically-limited model} \cite{7},\cite{9} which suppresses all effects due to
the diffusion. More precisely, we start from a {\em diffusion-limited model}
\cite{7}, in which the gel fractality results from the  brownian motion of the
particles which stick together with a probability $c$ when happening to occupy
nearest-neighbour lattice sites. Making then $c$ going to $0$, we end at the
chemical model mentioned above : the particle random path lenght before
aggregation goes to infinity, with the consequence that {\em all the free sites
on the aggregate have an equal probability of stick to it}, except for steric
hindrance. Of course the fractal dimensions obtained are superior to the
diffusion-limited model ones.\\
In our algorithm, the folowing are associated with every particle at every step
: (1) the list of its six sites and their state (hydrolyzed, non-hydrolyzed, or
already used in condensation); (2) the reference number of the cluster to which
it belongs; (3) its 3D coordinates {\em inside its cluster} (i.e. its position
relative to the other particles in the same cluster only). So each cluster is
on a lattice but the whole process is not, at least not in a box\cite{10}.\\
At every step, two bonds are selected randomly among the $6N$ ones. In order
for the aggregation process to take place, they must obey the following
conditions: (P1) they both must be hydrolyzed sites; (P2) they must belong to
two {\em different} particles {\em not being in the same cluster}
; (P3) moreover, cluster rotation is excluded, so the two bonds must be on two
opposites sites on the same symmetry axis of the monomer. Only if these
conditions are fulfilled can one then check the steric possibility of the
aggregation, due to the presence of other molecules already connected to the
selected sites.\\
Why this model? To neglect the hazards of the diffusion is equivalent to
accelerating the aggregation dynamics; this is an important advantage since we
wish to explore a two dimensional parameter space, which requires many
drawings. Although the physical polymerization process in our solutions is
conceivably controlled in part by the diffusion, its effect is only to globally
decrease the fractal dimension of the final aggregate, without modifying the
relative variations due to polydispersity effects. So the same trends must be
observed in this model for the dimension in the  $p$ and $q$ phase space,
unless other assumptions are made on the influence of surfactants during the
polymerization process, as discussed below.\\
Being not in a box, it was not possible to introduce a gelation test (like the
existence of a percolation cluster) giving to the algorithm a systematic
stopping condition. The end of the simulation happens when (1) all the
particles are in one cluster (2) $(6N)^2$ {\em successive  }attempts to select
bonds  fulfiling the (P1)-(P3) conditions have been unsucessful (3) $N^2$ {\em
successive  }sticking attempts have failed due to steric reasons. None of these
conditions, even the first, has anything to do with a physical gelation. One
can obtain the fraction $a_N$ of the initial N particles sample found in the
biggest final aggregate; to impose $a_N=1$ as a gelation criterion give a phase
transition diagram in the $p$, $q$ plane with a gelation domain narrowly
limited around the $q=1$ limit (value corresponding to total hydrolysis). This
is clearly not correlated to the {\em physical} gelation domain, which contain
all the $h>2$ part of the phase space. Moreover it is obvious that the physical
gelation does not necessitate {\em all} the monomers to be polymerized.\\
After the conclusion of the aggregation process, we compute the fractal
dimension of the biggest aggregate created. We obtain it from the double
logarithm graph of the two points correlation function calculated with the
following algorithm : for every particle the {\em array of distances of the
other monomers}, by increasing order, is obtained; one then averages over all
the arrays (i.e. over all the cluster monomers). With $c(r)$ the two-points
correlation function, and $D$ the fractal dimension, one has:
$$
n_R(D)=\int_0^R c(r) d^3r = R^D.
$$
And from the slope of the linear function
$$
log N = {\bf f}\Bigl( log R\Bigr)
$$
the fractal dimension is easily obtained.\\
\section{DISCUSSION}
With this method, we have done numerical experiments to be compared with Fig.
1, $q$ varying from $0.5$ to $0.75$  and $p$ from $0$ to $1$. The step for $q$
was $5 \times$ $10^{-2}$, and $10^{-1}$ for $p$. Due to the very nature of the
simulation algorithm, the cluster self-similarity is obtained only on distances
of approximately one-quarter of its maximum radius. Hence, to observe the
linearity of the correlation function on many orders of magnitude, it should be
necessary to work on particle samples of at least $10^4$ elements. On a RISK
workstation, the aggregation algorithm could take many hours to complete.
Nevertheless, it is remarkable that on $1000$ particles samples, sufficient
information can be obtained on the fractal dimension, a fact that has been
observed in all the ($p,q$)-phase space. This is all the more true if one
averages over many samples, a procedure which increases only {\em linearly} the
execution time. Nevertheless the dimensions are a little smaller (of $0.2$
approximately) than the physical ones.
 Fig. ~\ref{fig2} \/shows the proportion evolutions with p ("polydispersity")
of the different species for $q=0.5$ $(h=2)$ and  $q=0.75$ $(h=3)$. For the
same two extremal values of $q$, Fig. ~\ref{fig3} shows the fractal dimension
evolution with $p$. To compare with the experimental evolutions, one must
remember the correspondence $q \rightarrow h$ and $p \rightarrow n$. The $h=2$
plot of ~\ref{fig3} can be compared to the bottom ~\ref{fig1} experimental
plot. The $p$ variation at constant $h$ corresponds moreover to {\em vertical}
evolutions between the two plots of ~\ref{fig1} (top). So we predict that the
two evolutions with $h$ will {\em converge} for values of $q$ near 1, and not
{\em cross}, as in the experimental data.\\
For all the $q$ values, the fractal dimension increases with p. This is easily
understood from fig. ~\ref{fig2}. When $p=0$, the distribution of differently
hydrolyzed species shows no polydispersity : it is peaked on a value depending
on $q$; for example for $q=0.5$, {\em all} the particles have two hydrolyzed
sites , for  $q=0.75$ three hydrolyzed sites. As $p$ increases, the proportion
of particles having more hydrolyzed sites than this initial value increases,
and forms a greater proportion of the final aggregate (remember that this
cluster contains not necessarily all the particles of the sample). At the $p=1$
limit, only four-hydrolyzed-site particles remain.\\
The  important point is that when $q$ increases, the {\em domain of variation}
of the fractal dimension {\em decreases }with $p$. At the $q=1$ limit (complete
hydrolysis) of course the polydispersity cannot play any role, and $p$ has no
influence on the fractal dimension, since all the possible sites are
hydrolyzed. The asymptotic value of the dimension is then the same as for
$p=1$, since as previously said only particles with four hydrolyzed sites
remain for this value. Now for the evolution with $q$ of the dimension, as
previously mentioned, one of the difficulties of this simulation is the gauging
of the $n$ and $p$ scales. However, the increasing of the fractal dimension
with $h$ for $n=1$ can clearly be associated with the same evolution with $q$
for $p=0$. The ranges of evolution coincide rather well (Fig. ~\ref{fig3}).\\
For $n=2$, taking into account the error bars in the x-ray determination of the
gel structure, one could interpret in at least two ways the evolution of the
fractal dimension. The first is a {\em  convergence} of the $n=1$ and $n=2$
curves to a common asymptotic value ($2.4$), around which the $n=2$ curve would
not vary very much. This is in fact the prediction of our simulation for a pure
hydrolysis effect. For $p=1$, the dimension is constantly equal to the
asymptotic limit $2.35$. Taking for $n=2$ values of $p$ near 1 would give a
slight increase of the dimension with $h$.\\
However, the experimental evolution of the fractal dimension with $h$ for large
values of $n$ seemed to point out another interpretation, and could
definitively not be explained by our model. For $n=2$, due to one experimental
point, the fractal dimension {\em decreased with h}, and crossed the increasing
$n=1$ curve. No mechanism based on the {\em initial hydrolysis} could account
for this trend, since, given any kind of initial distribution of hydroxyl
radicals, the fractal dimension must always {\em increase} with the hydrolysis
rate, or at least stay constant if it appears that the hydrolysis has no effect
on the structure of the gel. In any case, for $h \rightarrow 4$, the dimension
must for any value of $n$ converge to the same limit, correponding to the {\em
total hydrolysis case}, for which the distribution of the radicals has no role
to play. The lack of experimental points in this region of the $n, h$ phase
diagram led us to perform a second set of measures at $n=2$ and $2.5<h<3$. With
more values, the fractal dimension appears in this range remarkably constant
(see fig. ~\ref{fig4}), and coherent with the predicted asymptotic value
$2.35$. We concluded that the $n=2, h=3$ value of fig. ~\ref{fig1} was not
reproductible. The evolution in the ($h,n$) phase space predicted by the model
seems to be confirmed.  \\
\section{CONCLUSION}
We have succeeded in understanding through simulation the fractal dimension
evolution for all the regions of the ($h,n$) phase diagram, with a  hydrolysis
polydispersity hypothesis in the first stage of the reaction. The dimension
increases with $h$ and $n$ in a well understood way. Therefore it appears that
the influence of the surfactants molecules on the sol-gel process could be
limited to the first step of the gelification. Interactions occcuring later, in
particular with the hydroxyl radicals, seem to be without effect on the
structural properties of the gel. \\

Aknowledgments. We wish to thank J. Lambard, P. Lesieur, and T. Zemb, who
allowed us to realize ultra small angle x-scattering experiments and data
treatments in the "Service de Chimie Mol\'eculaire" at the C.E.N., Saclay.

\begin{figure}
\caption{Experiments : fractal dimension variation with $h$ for $n=1$ and $n=2$
(bottom); same variation with $n$ for $h=2$ (top).}
\label{fig1}
\end{figure}

\begin{figure}
\caption{Polydispersity variation with $p$ for $q=0.75$ (top) and $q=0.5$
(bottom).}
\label{fig2}
\end{figure}

\begin{figure}
\caption{Simulation : fractal dimension variation with $p$ for $h=2$ and
$h=3$..}
\label{fig3}
\end{figure}

\begin{figure}
\caption{Experiment : evolution of the dimension for $n=2$, $2.5<h<3$.}
\label{fig4}
\end{figure}

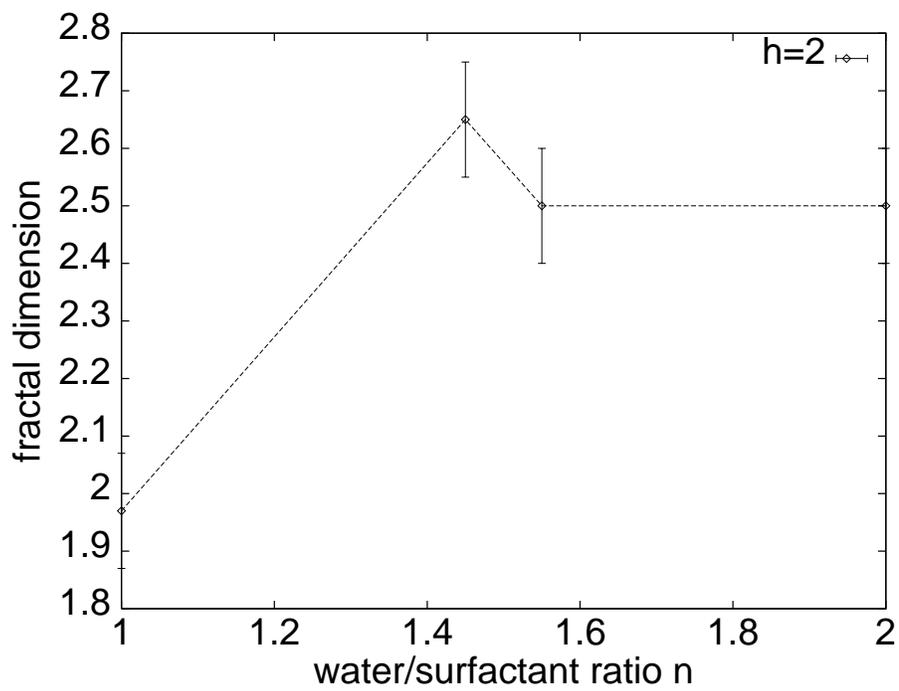
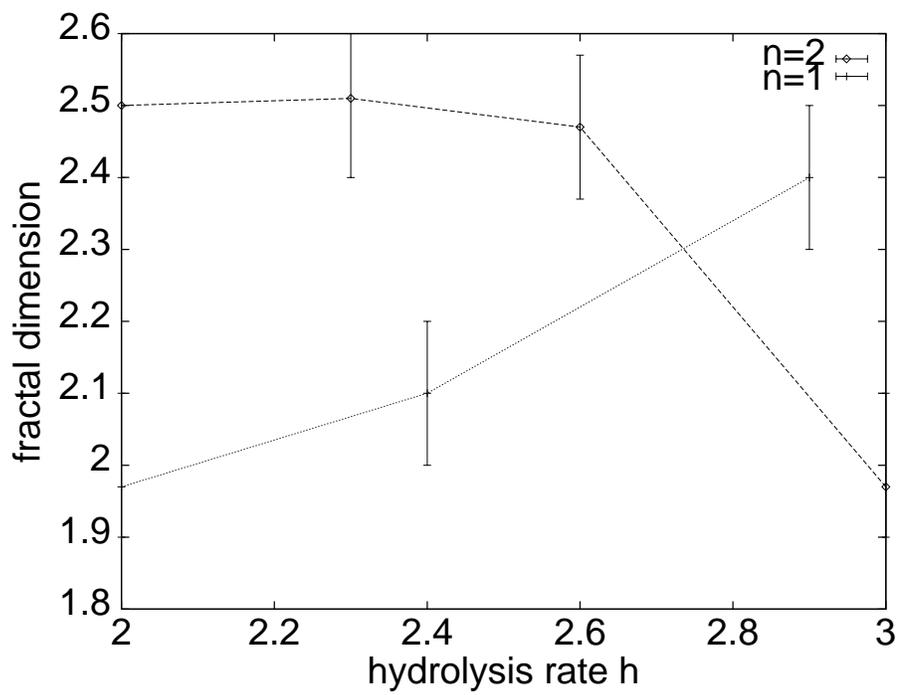
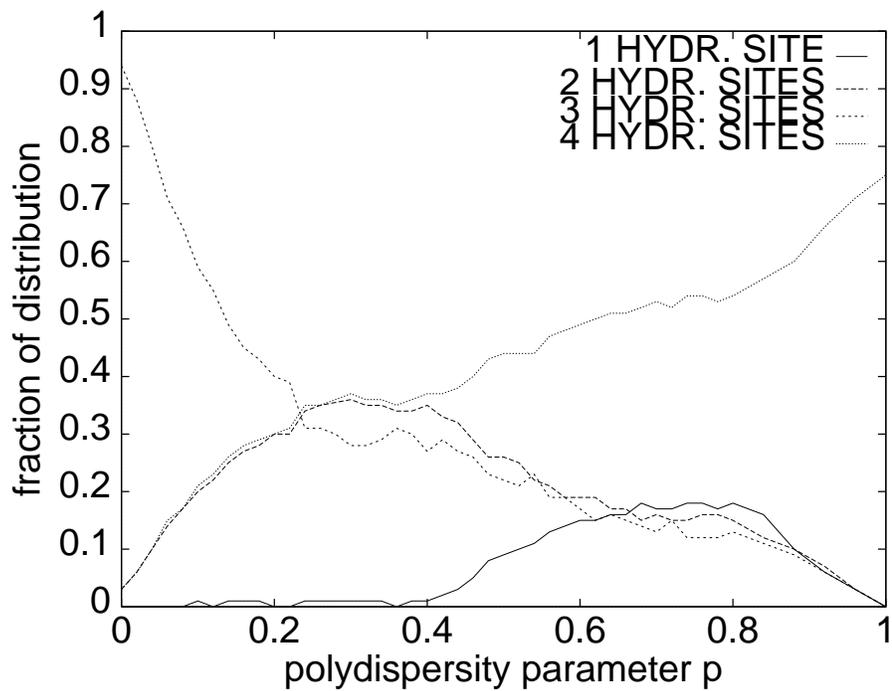
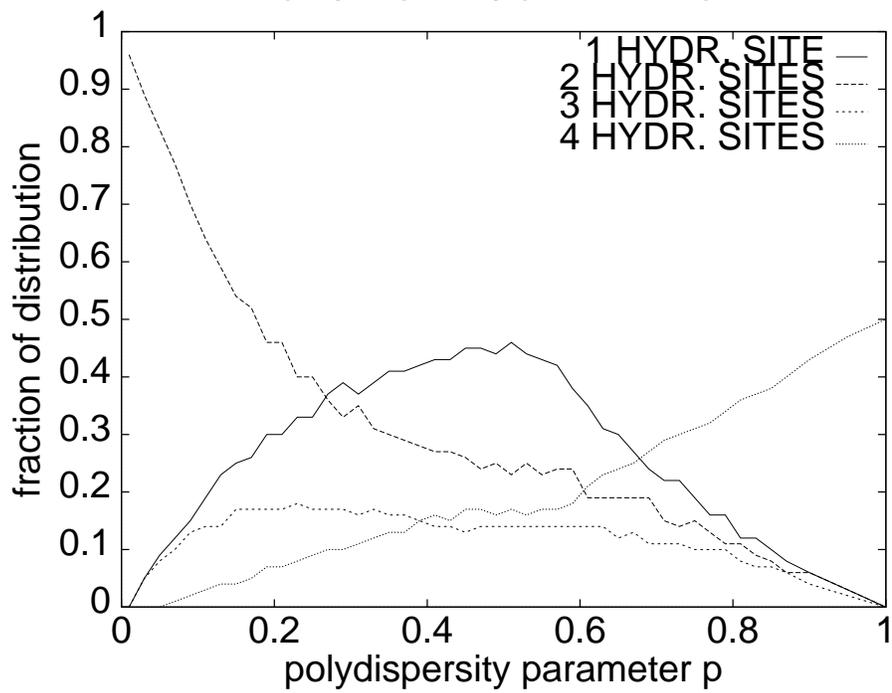
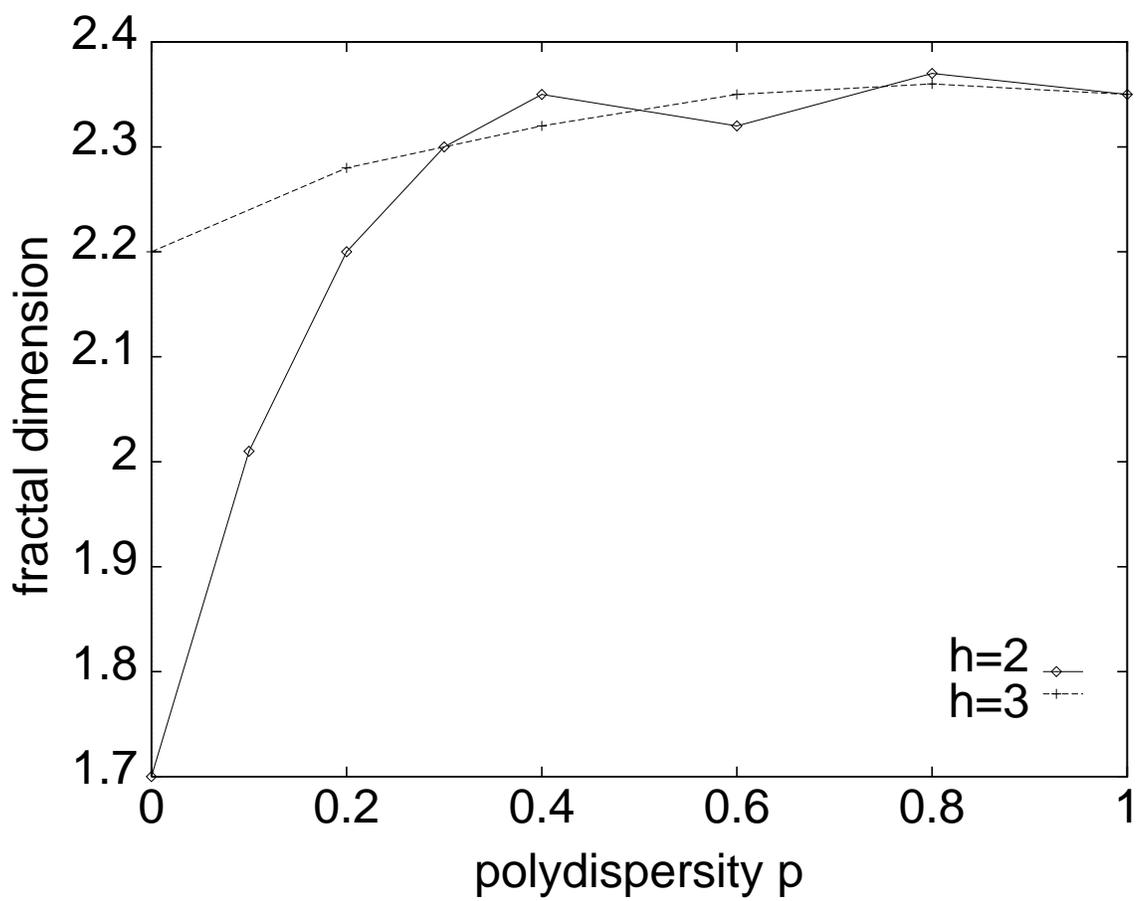
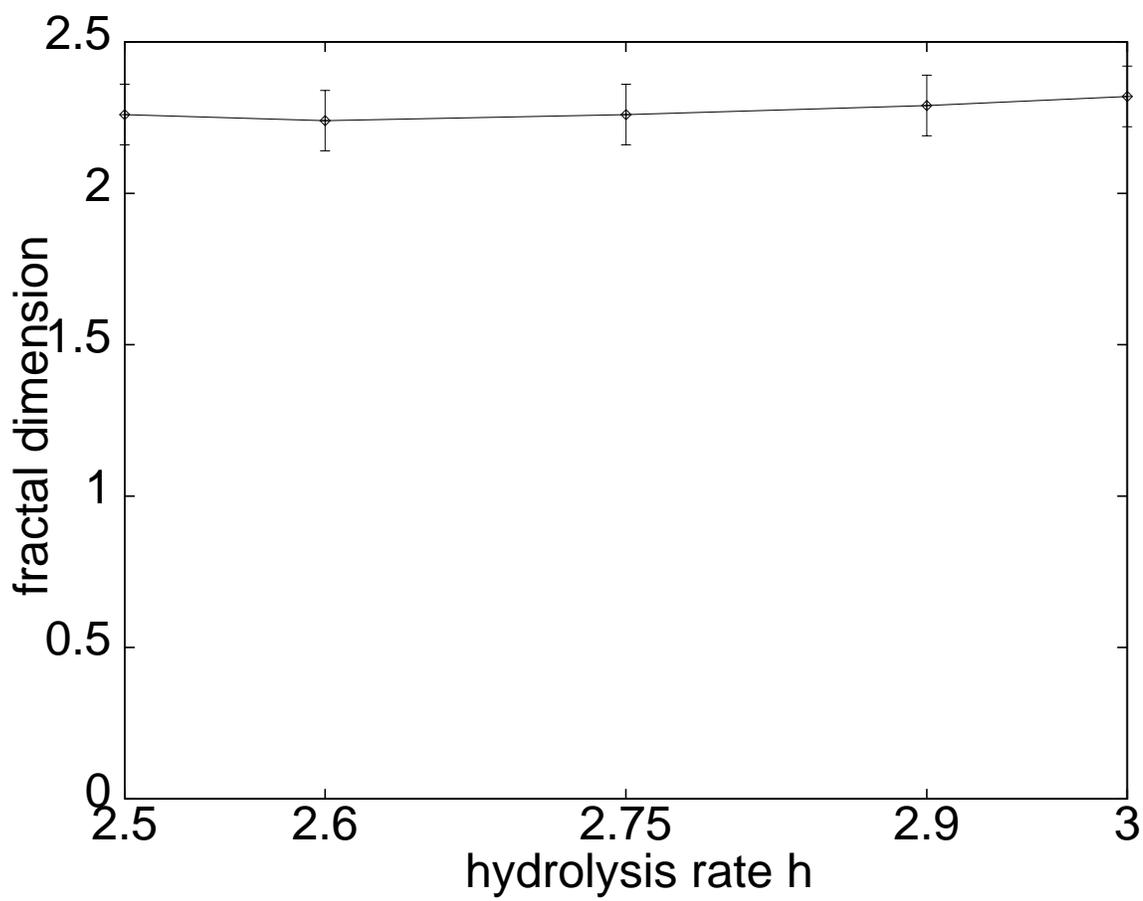
\end{document}